\documentclass[twocolumn]{aastex631}

\newcommand\target{{Cyg X-2}}
\newcommand\nicer{{\it NICER}}

\begin{document}

\title{Fast Transitions of X-ray Variability in the Neutron Star Low Mass X-ray Binary Cygnus X-2}

\correspondingauthor{Liang Zhang}
\email{zhangliang@ihep.ac.cn}

\author[0000-0002-0786-7307]{Liang Zhang}
\affiliation{Key Laboratory of Particle Astrophysics, Institute of High Energy Physics, Chinese Academy of Sciences, Beijing 100049, People's Republic of China}

\author{Mariano M\'{e}ndez}
\affiliation{Kapteyn Astronomical Institute, University of Groningen, P.O. Box 800, NL-9700 AV Groningen, The Netherlands}

\author{Hua Feng}
\affiliation{Key Laboratory of Particle Astrophysics, Institute of High Energy Physics, Chinese Academy of Sciences, Beijing 100049, People's Republic of China}

\author{Diego Altamirano}
\affiliation{School of Physics and Astronomy, University of Southampton, Southampton, Hampshire SO17 1BJ, UK}

\author{Zi-xu Yang}
\affiliation{School of Physics and Optoelectronic Engineering, Shandong University of Technology, Zibo 255000, People's Republic of China}

\author{Qing-chang Zhao}
\affiliation{Key Laboratory of Particle Astrophysics, Institute of High Energy Physics, Chinese Academy of Sciences, Beijing 100049, People's Republic of China}

\author[0000-0001-5586-1017]{Shuang-Nan Zhang}
\affiliation{Key Laboratory of Particle Astrophysics, Institute of High Energy Physics, Chinese Academy of Sciences, Beijing 100049, People's Republic of China}

\author[0000-0002-2705-4338]{Lian Tao}
\affiliation{Key Laboratory of Particle Astrophysics, Institute of High Energy Physics, Chinese Academy of Sciences, Beijing 100049, People's Republic of China}

\author[0000-0002-3515-9500]{Yue Huang}
\affiliation{Key Laboratory of Particle Astrophysics, Institute of High Energy Physics, Chinese Academy of Sciences, Beijing 100049, People's Republic of China}

\author{Xiang Ma}
\affiliation{Key Laboratory of Particle Astrophysics, Institute of High Energy Physics, Chinese Academy of Sciences, Beijing 100049, People's Republic of China}

\author{Shu-mei Jia}
\affiliation{Key Laboratory of Particle Astrophysics, Institute of High Energy Physics, Chinese Academy of Sciences, Beijing 100049, People's Republic of China}

\author{Ming-yu Ge}
\affiliation{Key Laboratory of Particle Astrophysics, Institute of High Energy Physics, Chinese Academy of Sciences, Beijing 100049, People's Republic of China}

\author[0000-0003-0274-3396]{Li-Ming Song}
\affiliation{Key Laboratory of Particle Astrophysics, Institute of High Energy Physics, Chinese Academy of Sciences, Beijing 100049, People's Republic of China}

\author[0000-0002-9796-2585]{Jin-Lu Qu}
\affiliation{Key Laboratory of Particle Astrophysics, Institute of High Energy Physics, Chinese Academy of Sciences, Beijing 100049, People's Republic of China}

\author{Shu Zhang}
\affiliation{Key Laboratory of Particle Astrophysics, Institute of High Energy Physics, Chinese Academy of Sciences, Beijing 100049, People's Republic of China}



\begin{abstract}
We present a spectral-timing analysis of two \nicer\ observations of the weakly magnetized neutron star low-mass X-ray binary Cygnus X-2. 
During these observations, we detect a rapid transition from a narrow 50-Hz horizontal-branch oscillation to a broad 5-Hz normal-branch oscillation, accompanied by an increase in source flux and a decrease in spectral hardness.
Thanks to the large effective area of \nicer, we are able to conduct a detailed comparison of the spectra associated with different types of quasi-periodic oscillations (QPOs) on short timescales.
By fitting the spectra with a model that includes a disc and Comptonization components plus two emission lines, we find that the parameters of the disc component do not change significantly during the transition. However, assuming a fixed electron temperature, the optical depth of the Comptonization component decreases significantly. This drop in optical depth may be attributed to the expansion of the boundary layer or spreading layer.
In addition, we find that the rms spectra for both the HBO and NBO are hard, suggesting that the boundary layer or spreading layer is driving the variability. 
We discuss the potential physical origin of the different types of QPOs.

\end{abstract}

\keywords{Low-mass X-ray binary stars (939); Accretion (14); Neutron stars (1108); X-ray astronomy (1810)}

\section{Introduction} 
\label{sec:intro}

Weakly magnetized neutron stars (NSs) in low-mass X-ray binaries (LMXBs) accrete matter from a low-mass companion star via Roche lobe overflow \citep[see][for a recent review]{Bahramian2023}.
The infalling matter forms an accretion disc around the compact object, with the gas rotating approximately in Keplerian orbits \citep{Shakura1973}. As the accreting gas approaches the surface of the NS, half of the gravitational potential energy released in the accretion process is dissipated in a boundary layer (BL, \citealt{Popham2001}).
Meanwhile, the spreading of matter over the surface of the NS creates a spreading layer (SL, \citealt{Inogamov1999}).

Based on their tracks in the X-ray color-color diagram (CCD) or hardness-intensity diagram (HID), NS-LMXBs can be categorized into Z- and atoll sources \citep{Hasinger1989}.
Z sources, which typically exhibit higher luminosity (close to the Eddington limit) compared to atoll sources, trace out a distinct Z-shaped path in the CCD/HID on timescales of hours to days.
This distinctive Z-like track is characterized by three branches: horizontal branch (HB), normal branch (NB), and flaring branch (FB).

QPOs are common features in accreting systems. They appear in the power density spectrum (PDS) as narrow peaks, often accompanied by different forms of broadband noise components \citep[see][for a recent review]{Ingram2019}. 
In NS-LMXBs, QPOs have been observed across a broad range of frequencies. These QPOs can be categorized into three main groups based on their specific frequency ranges: millihertz QPOs \citep{Revnivtsev2001,Altamirano2008,Lyu2015,Mancuso2023}, low-frequency QPOs \citep{Wijnands1999,Homan2002,Homan2007,Altamirano2012,Motta2017}, and kilohertz QPOs (see \citealt{Mendez2021} and references therein).
In Z sources, low-frequency QPOs can be further differentiated based on the specific branch occupied by the source in the CCD/HID at the time the QPO is observed. Three main types of low-frequency QPOs have been identified: horizontal-branch oscillations (HBOs), normal-branch oscillations (NBOs), and flaring-branch oscillations (FBOs) \citep{Klis1989}.
The frequency of the HBO typically ranges from 10 to 60 Hz, while NBO and FBO are usually observed at frequencies around 5$-$7 Hz and 15 Hz, respectively \citep{Motta2017}.
Despite numerous efforts to characterize these phenomena, there is currently no consensus on the physical origins of the various types of QPOs \citep[e.g.,][]{Fortner1989,Alpar1992,Stella1998,Titarchuk2001,Ingram2010}.

Fast transitions between different types of low-frequency QPOs have been observed in several black hole (BH) LMXBs \citep[e.g.,][]{Homan2001,Soleri2008,Homan2020,Sriram2021,Zhang2021,Yang2023}. Studying these transitions not only provides crucial evidence regarding the physical origin of the QPOs but also enhances our understanding of the mechanisms driving state transitions.
\citet{Soleri2008} investigated the transition from the so-called type-C QPO to the type-B QPO in the black hole LMXB GRS 1915+105 and suggested that the transition is related to discrete jet ejection. \citet{Ma2023} reported the transition from the type-C to the type-B QPO in MAXI J1820+070. Their study revealed a significant change in the geometry of the corona during the transition, shifting from a horizontally extended configuration to a vertically extended one.
In contrast, fast QPO transitions in NS-LMXBs have received limited attention in the past.

Cygnus X-2 (\target) is a luminous and persistently active LMXB that is classified as a Z source based on the characteristic shape of its tracks in the CCD/HID \citep{Hasinger1990}.
The mass of the NS in \target\ is estimated to be $1.71 \pm 0.21 M_{\odot}$ for an inclination of $62.5^{\circ} \pm 4^{\circ}$ \citep{Orosz1999,Casares2010}. Its stellar companion is a late-type A9III star with an orbital period of $\sim$9.8\,d \citep{Casares1998}. The system is located at a distance estimated to be between 7 and 11 kpc \citep{Orosz1999,Ding2021}.

The broadband X-ray spectra of \target\ have been studied extensively. 
\citet{Disalvo2002} found that the {\it BeppoSAX} spectra of \target\ can be described by a soft multicolor component from the accretion disc and a Comptonization component likely originating from the hot BL/SL.
Moreover, the spectra revealed the presence of a reflection component in the form of a broadened Fe K line at around 6.4 keV, as well as an emission line near 1 keV \citep{Done2002,Disalvo2002,Farinelli2009,Ludlam2022}.
%
%
HBO at 10--60 Hz and NBO at around 5 Hz have been detected in the PDS of \target\ \citep{Wijnads2001,Motta2017,Chhangte2022,Jia2023,Sudha2025}.

In this work, we report the detection of a fast transition from a HBO to a NBO in two \nicer\ observations of \target. 
We describe our observations and data reduction methodology in Section \ref{sec:data}. We present our spectral-timing results in Section \ref{sec:results} and discuss them in Section \ref{sec:discussion}.

\section{Observations and data analysis} \label{sec:data}

\subsection{Data selection}

\begin{table}
\caption{Log of the two \nicer\ observations of \target\ in which a fast transition between different types of QPOs was observed.}
\label{tab:log}
\begin{center}
\begin{tabular}{lccc}
\hline
\hline
Obs & Obs. ID & Obs.\ Start Date & Exposure (ks) \\
\hline
\#1 & 1034150111 & 2017-10-21 23:34:20 &  24.6 \\
\#2 & 1034150112 & 2017-10-23 00:13:48 &  14.6 \\

\hline

\end{tabular}
\end{center}
\end{table}

\nicer\ is a soft X-ray telescope onboard the International Space Station (ISS; \citealt{Gendreau2016}). \nicer\ provides high throughput in the 0.2--12 keV energy band,  with a large effective area of $\sim$1900 $\rm{cm}^2$ at 1.5 keV and an exceptional absolute timing precision of around 100 ns. These properties make it an ideal instrument for studying fast X-ray variability.

We consider all the \nicer\ archival observations of \target\ spanning from 2017 June 25 to 2024 August 31.
All data were reprocessed using the \nicer\ software tools NICERDAS version 2024-02-09\_V012A, along with the calibration database (CALDB) version xti20240206. 
We filtered the data with the standard screening criteria via the \texttt{nicerl2} task. 
Additionally, we divided the good time intervals of each observation into multiple continuous data segments based on the orbit of the ISS for further analysis.

To identify rapid changes in X-ray variability, we produced an average PDS for each data segment in the 0.5--10 keV energy band. 
We used a time interval of 16\,s and a time resolution of 1/8192\,s.
The resulting PDS were normalized in units of (rms/mean)$^{2}$~Hz$^{-1}$ \citep{Belloni1990}, and the Poisson noise level estimated from the power between 3000 and 4000 Hz was subtracted. 
Upon inspection, we detected fast transitions between different types of QPOs in two observations. The observation IDs, dates, and net exposure times are listed in Table \ref{tab:log}. In the remainder of this paper, we will focus our analysis on the characteristics of these two observations.

\subsection{Timing and spectral analysis}

For each of the two observations, we extracted light and hardness ratio curves with a binning of 80\,s in different energy bands using the \texttt{nicerl3-lc} routine.
%
%
%
The light curves were not background subtracted, as the background contribution (typically less than 2 counts s$^{-1}$) is negligible compared to the source count rate (3500--8500 counts s$^{-1}$ during the observations we analyzed).

The dynamical and average PDS shown in this paper were calculated in the 0.5--10 keV as previously described.
We fitted the PDS using a model composed of a combination of Lorentzian functions.

We used the \texttt{nicerl3-spect} routine to extract both the total and background spectra, as well as the response files. The 3C50 background model was employed for spectral analysis \citep{Remillard2022}.
We binned the spectra following the optimal binning algorithm proposed by \citet{Kaastra2016} with a minimum of 30 counts per energy bin. Subsequently, we performed spectral fitting in the 0.5–-10 keV range using XSPEC version 12.14.0.

\section{Results}
\label{sec:results}

\subsection{Fast transition between different types of QPOs}

\begin{figure*}[ht]
\begin{center}
\resizebox{2\columnwidth}{!}{\rotatebox{0}{\includegraphics{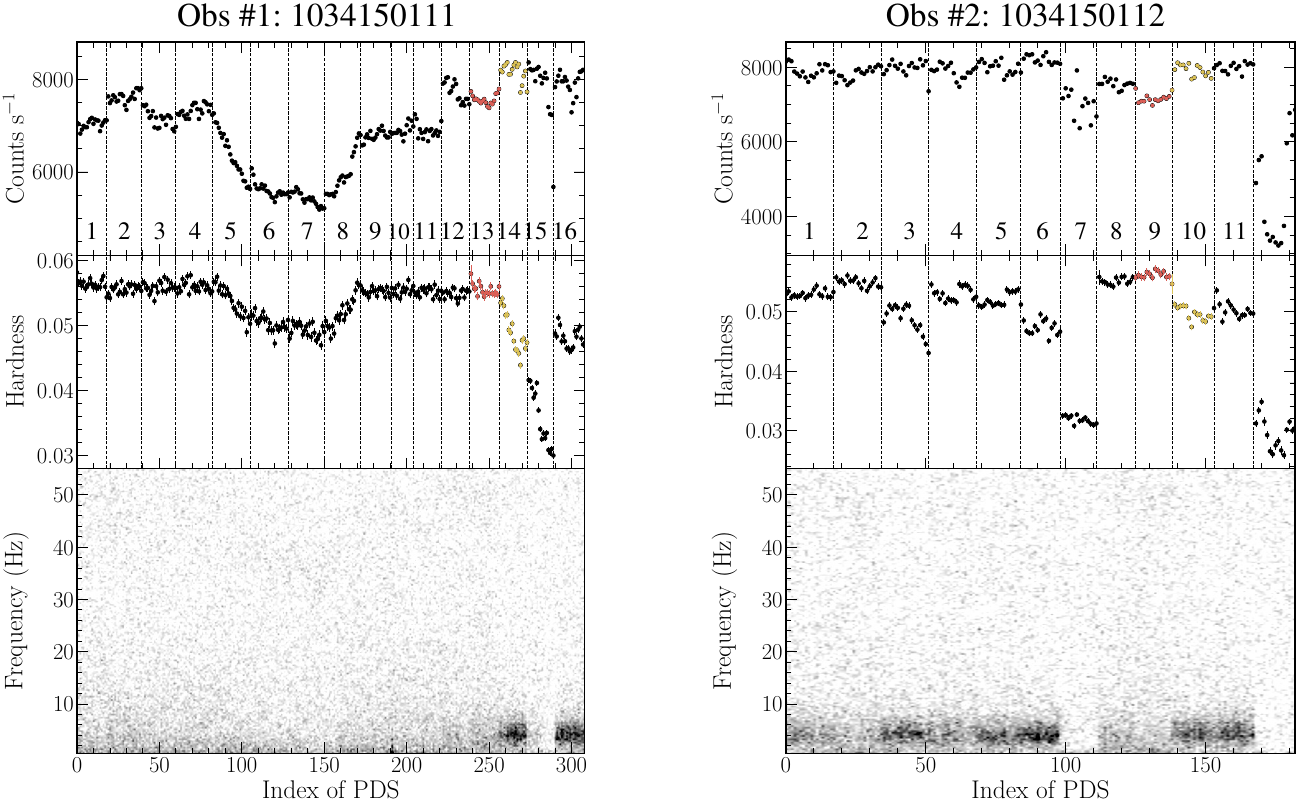}}}
\end{center}
\caption{Detailed look at the two \nicer\ observations of \target\ in which a fast transition between different types of QPOs was detected.
The top and middle panels show the evolution of the 0.5--10 keV count rate and hardness ratio (6--10 keV/2--4 keV) with a time resolution of 80\,s. The bottom panels show the dynamical power spectra. The PDS were initially calculated with a time interval of 16\,s and subsequently rebinned by a factor of 5 both in time and frequency.
All time gaps were removed and marked with dotted lines. The $x$-axis represents the index of each PDS. For the dynamical power spectra, the powers were not rms-normalized and the Poisson noise was not subtracted.
The red (yellow) points mark the orbit before (after) the transition.}
\label{fig:dpds}
\end{figure*}

In Figure \ref{fig:dpds}, we show the evolution of the 0.5--10 keV count rate and hardness ratio (6--10 keV/2--4 keV), together with the corresponding dynamical PDS for the two observations we analyzed. All time gaps have been removed and are marked with dotted lines.
In Obs \#1, a significant QPO around 5 Hz is visible in the dynamical PDS of orbit 14. However, this QPO is absent in orbit 13, where broadband noise can be seen at low frequencies. In orbit 15, neither the 5-Hz QPO nor the broadband noise is clearly observed, while the 5-Hz QPO reappears in orbit 16.
This paper focuses on the transition between orbits 13 and 14, during which we observed rapid shifts between different types of QPOs (detailed in the following sections). The transition between orbits 14 and 15, associated with the disappearance of the QPO, has been investigated by \citet{Sudha2025} and is not the focus of our study.
A similar fast transition was also identified in Obs \#2, occurring between orbits 9 and 10. A 5-Hz QPO appeared in orbit 10 but was undetectable in orbit 9.
Remarkably, the orbits exhibiting the 5-Hz QPO show a higher flux and a lower hardness ratio compared to the orbits without the 5-Hz QPO.
We note that since no fast transitions were detected within a single orbit, we can only constrain the transition timescale to be less than $\sim$1 hour, corresponding to the data gap between the two orbits.

\begin{figure*}[ht]
\begin{center}
\resizebox{2\columnwidth}{!}{\rotatebox{0}{\includegraphics{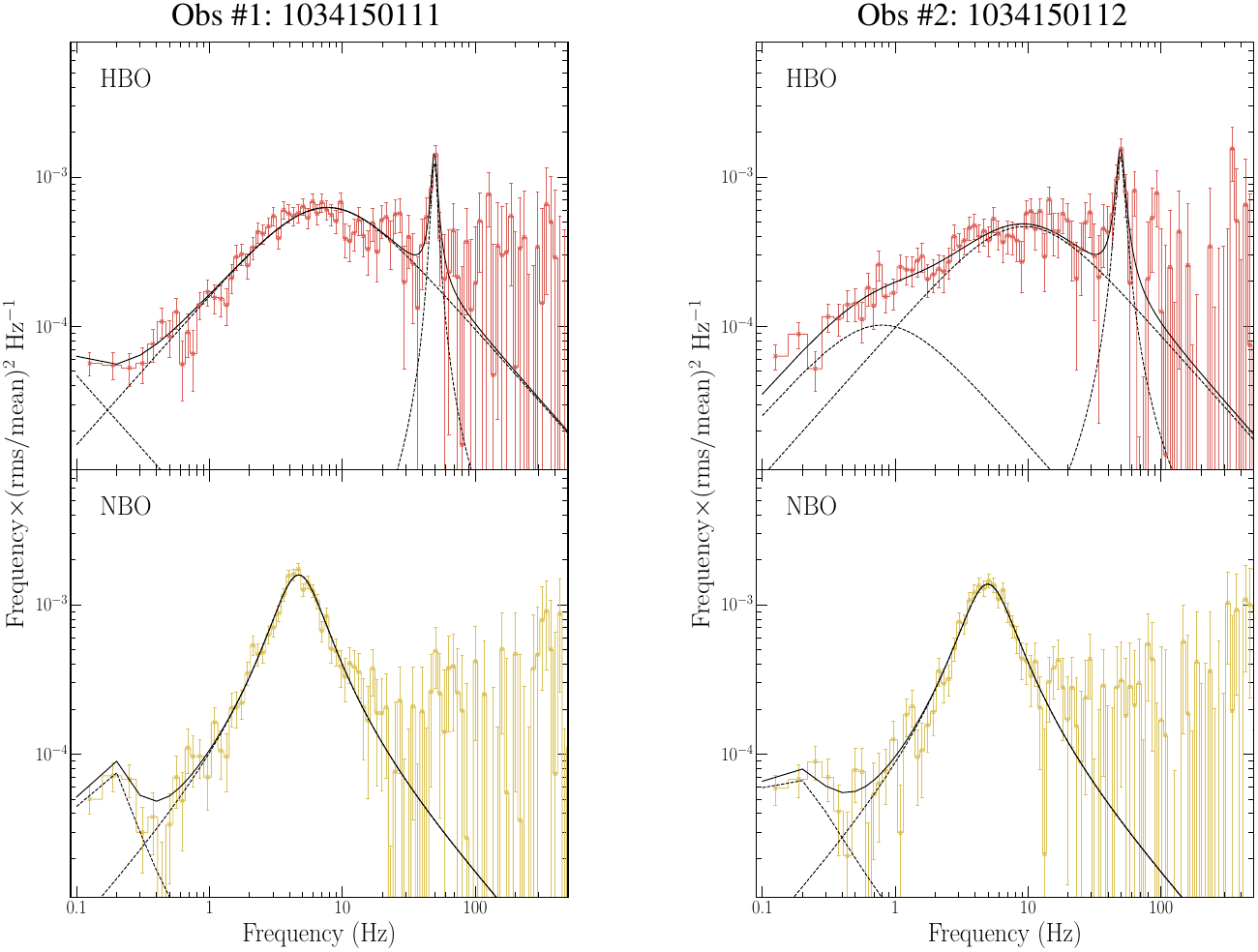}}}
\end{center}
\caption{Power spectra of \target\ averaged from the orbit before (red) and after (yellow) the rapid transition for the two \nicer\ observations. The average power spectra were calculated in the 0.5--10 keV energy band and rms-normalized, with the contribution due to Poisson noise subtracted. The power spectra were fitted with a model composed of a combination of Lorentzian functions. Before the transition, a $\sim$50 Hz HBO was detected, while a $\sim$5 Hz NBO was observed after the transition.}
\label{fig:pds}
\end{figure*}

\begin{figure}[ht]
\begin{center}
\resizebox{\columnwidth}{!}{\rotatebox{0}{\includegraphics{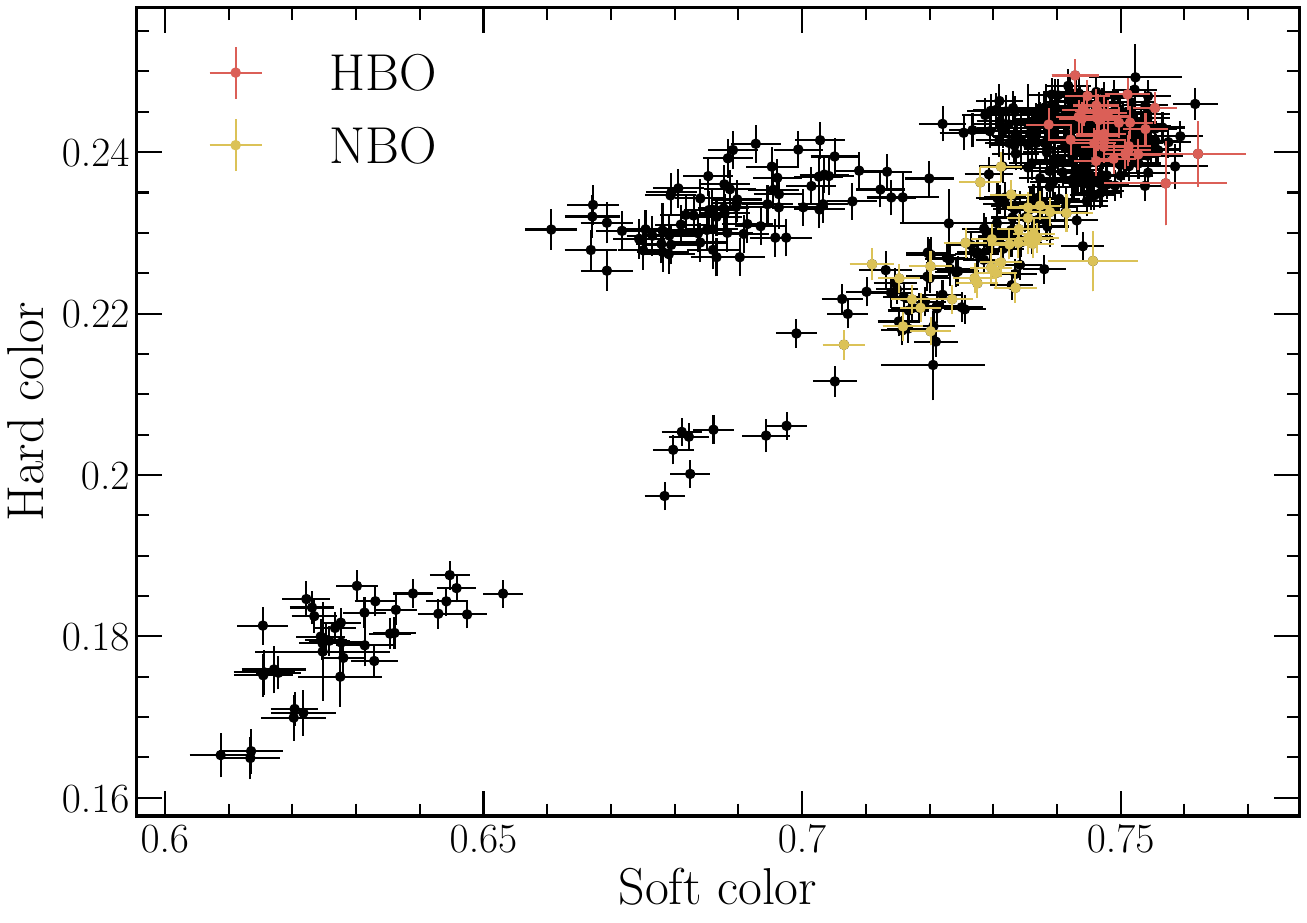}}}
\end{center}
\caption{Color-color diagram of \target\ for the two \nicer\ observations analyzed in this work. Hard color is defined as the ratio of count rates in the 5--8 keV to those in the 3--5 keV band, while soft color is defined as the ratio of count rates in the 3--5 keV to those in the 2--3 keV band. The orbits preceding the transition, characterized by the presence of HBOs, are marked in red, whereas the orbits following the transition with NBOs are highlighted in yellow.}
\label{fig:ccd}
\end{figure}

\begin{table*}
    \centering
    \caption{Parameters of the HBO and NBO observed during the rapid transition for the two \nicer\ observations of \target\ that we analyzed. The PDS were extracted from the single orbit before and after the transition.}
    \label{tab:qpo_para}
    \begin{tabular}{lcccccccc}
    \hline\hline
        Obs  &  Obs. ID    &   \multicolumn{3}{c}{HBO}   &&   \multicolumn{3}{c}{NBO} \\
         \cline{3-5} \cline{7-9} 
             &             &   $\nu_{0}$ (Hz) &   $Q$   &  rms (\%)  &&  $\nu_{0}$ (Hz) &   $Q$   &  rms (\%) \\
        \hline
        
        \#1  &  1034150111  & $49.1 \pm 0.7$  & $9.1 \pm 2.6$ & $1.6 \pm 0.2$ 
                            && $4.70 \pm 0.06$  & $1.03 \pm 0.05$ & $4.43 \pm 0.05$ \\
        
        \#2  &  1034150112  & $49.6 \pm 0.9$  & $6.2 \pm 2.3$ & $1.8 \pm 0.2$ 
                            && $4.98 \pm 0.08$  & $0.97 \pm 0.05$ & $4.20 \pm 0.06$ \\
       
        \hline
    \end{tabular}
\end{table*}

In Figure \ref{fig:pds}, we show the average PDS obtained from the orbits before (red) and after (yellow) the rapid transitions for the two observations we analyzed. 
In Figure \ref{fig:ccd}, we plot the two observations in the CCD. The data points corresponding to the orbits before and after the rapid transitions are marked in red and yellow, respectively.
In both observations, a sharp ($Q \gtrsim 6$) $\sim$50-Hz HBO was detected before the rapid transition, with a significance\footnote{Here the significance of QPOs is given as the ratio of the integral of the power of the Lorentzian used to fit the QPO divided by the negative 1$\sigma$ error on the integral of the power.} of around 4.5$\sigma$.
This QPO appears at the end of the HB and has a 0.5--10 keV fractional rms amplitude of $1.55 \pm 0.16$\% in obs \#1 and $1.83 \pm 0.20$\% in obs \#2. We note that this QPO is only significantly detected in the average PDS, without being clearly visible in the dynamical PDS.
In addition to the QPO, a peaked noise component was also present in the average PDS, fitting well with a Lorenzian function characterized by a frequency of approximately 8 Hz and a 0.5–10 keV fractional rms of $\sim$4\%.
After the transition, a broad ($Q\sim1$) $\sim$5-Hz NBO was observed in the PDS. This QPO emerges at the upper NB and has a 0.5--10 keV fractional rms of $4.43 \pm 0.05$\% in obs \#1 and $4.20 \pm 0.06$\% in obs \#2.
This 5-Hz NBO seems to evolve from the peaked noise observed in the PDS prior to the transition.
Notably, no significant broadband noise was detected alongside the 5-Hz NBO.
The PDS after the transition did not show a significant 50-Hz QPO, with a 3$\sigma$ upper limit for the 0.5--10 keV fractional rms amplitude of $\lesssim$1.2\%.
%
%
The main parameters of the HBO and NBO are listed in Table \ref{tab:qpo_para}.


\begin{figure}[ht]
\begin{center}
\resizebox{\columnwidth}{!}{\rotatebox{0}{\includegraphics{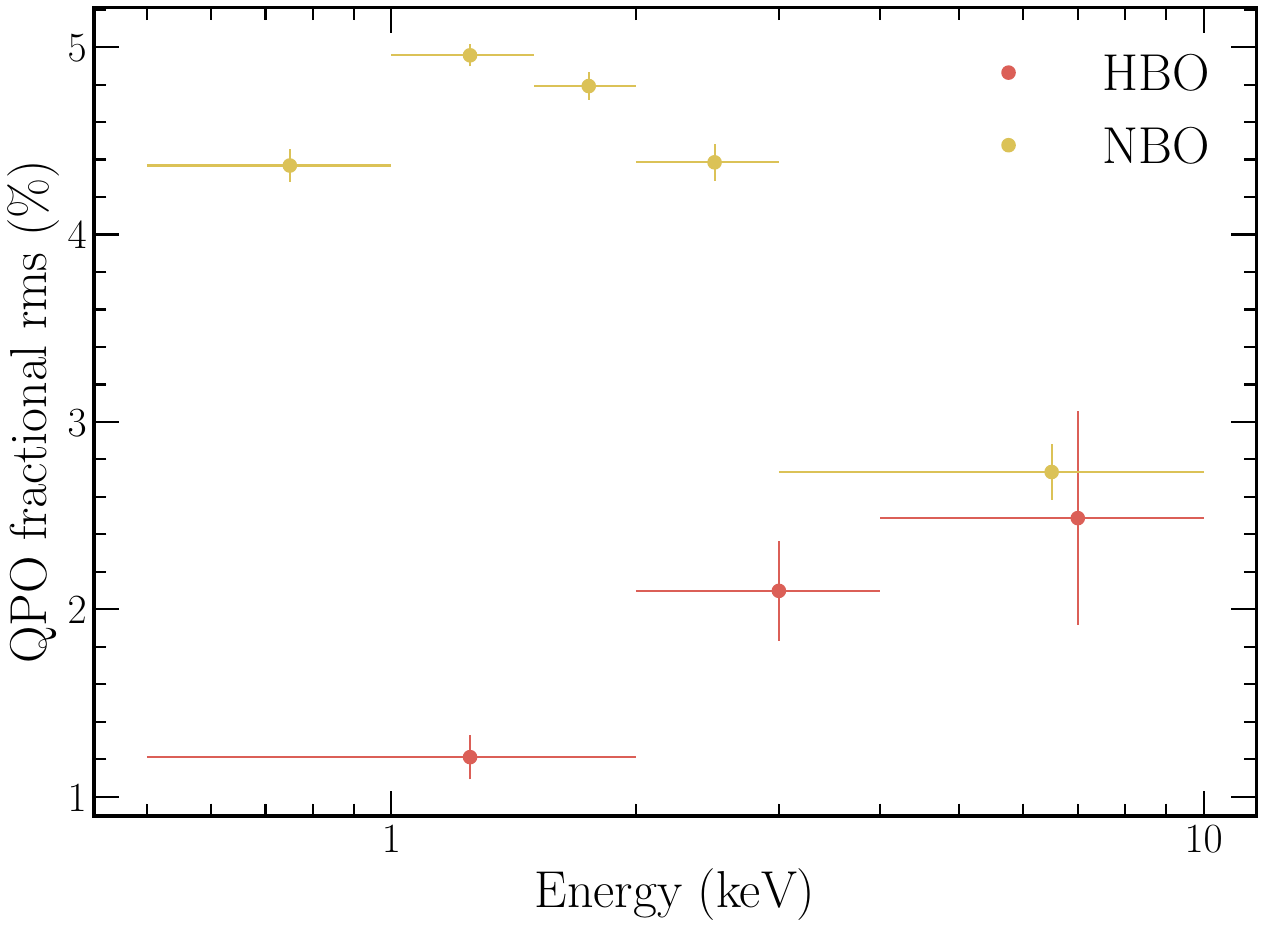}}}
\end{center}
\caption{Fractional rms amplitude of the HBO (red) and NBO (yellow) observed in \target\ as a function of photon energy.}
\label{fig:qpo_rms}
\end{figure}

We further investigated the energy-dependent fractional rms amplitude of the HBO and NBO by extracting PDS from different energy bands, following the same procedure as described earlier. 
In the two observations, the frequencies of, respectively, the HBO and the NBO are notably similar. Consequently, we merged the data segments corresponding to each type of QPO for our energy-dependent analysis.
Based on the available statistics, we calculated the fractional rms of the HBO in the energy bands of 0.5--2.0 keV, 2.0--4.0 keV, and 4.0--10.0 keV. For the NBO, we computed the fractional rms in the energy bands of 0.5--1.0 keV, 1.0--1.5 keV, 1.5--2.0 keV, 2.0--3.0 keV, and 3.0--10.0 keV. 
In Figure \ref{fig:qpo_rms}, we show the fractional rms of the QPOs as a function of photon energy.
We observed that the fractional rms of the NBO peaks around 1 keV before gradually decreasing at higher energies, while the fractional rms of the HBO exhibits a slight increase with photon energy.

\subsection{Spectral difference}

\begin{figure*}[ht]
\begin{center}
\resizebox{2\columnwidth}{!}{\rotatebox{0}{\includegraphics{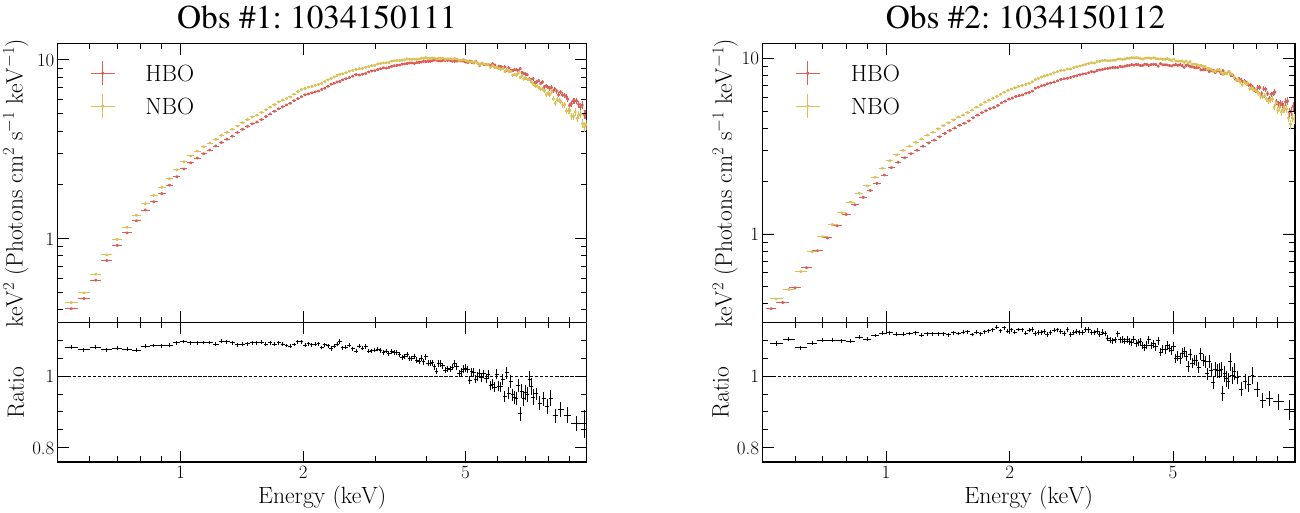}}}
\end{center}
\caption{Unfolded spectra of the orbit with a HBO (red) and a NBO (yellow), and corresponding spectral ratio (NBO/HBO) defined as the ratio between the spectrum with the NBO and that with the HBO for the two \nicer\ observations of \target\ that we analyzed. The spectra were deconvolved against a power law with $\Gamma = 2$.}
\label{fig:ratio}
\end{figure*}

To check potential spectral changes during the transition, we conducted spectral analysis by extracting spectra from the orbit with the HBO and the NBO separately for the two observations we analyzed. Hereafter, we will refer to these as the HBO and NBO spectra, respectively.
The HBO and NBO spectra, along with their corresponding spectral ratios (NBO/HBO), are shown in Figure \ref{fig:ratio}.
We found that the shapes of the spectral ratio between the two observations are remarkably similar. Below $\sim$3 keV, the spectral ratios are greater than unity and remain nearly constant. Above $\sim$3 keV, the spectral ratios gradually decrease to a value below unity, implying that the hard component is flatter in the HBO spectrum than in the NBO spectrum.

\subsection{Spectral modeling}

\begin{table*}

\centering
\caption{Best-fitting spectral parameters of the joint fit to the HBO and NBO spectra with the model \texttt{tbfeo*(diskbb+comptt+gaussian+gaussian)}. The fluxes of the different components were calculated in the $0.5-10$ keV band and are shown in units of $10^{-9}~\rm erg~cm^{-2}~s^{-1}$.}
\label{tab:spectral_fits}
\begin{tabular}{lcccccc}
\hline\hline
Component   &   Parameter   &   \multicolumn{2}{c}{1034150111}   &&  \multicolumn{2}{c}{1034150112} \\
 \cline{3-4} \cline{6-7} 
            &               &     HBO     &      NBO      &&     HBO     &      NBO  \\
\hline

TBFEO    &  $N_{\rm H}$ ($\times10^{22}~{\rm cm}^{-2}$)    &  \multicolumn{2}{c}{$0.24 \pm 0.01$} 
                                                           &&  \multicolumn{2}{c}{$0.24 \pm 0.01$} \\      
                                                      
DISKBB   &  $kT_{\rm in}$ (keV)    &  $0.59 \pm 0.02$  &  $0.60 \pm 0.02$    
                                   && $0.56 \pm 0.02$  &  $0.58 \pm 0.02$   \\

         &  Dnorm                   &  $3688 \pm 482$   &  $3802 \pm 501$
                                   && $4284 \pm 488$   &  $4260 \pm 502$    \\

COMPTT   &  $kT_{0}$ (keV)          &  $0.86 \pm 0.02$     &   $0.86 \pm 0.02$         
                                    && $0.83 \pm 0.01$     &   $0.85 \pm 0.01$   \\

         &  $kT_{\rm e}$ (keV)      &  \multicolumn{5}{c}{ [3] }  \\
         
         &  $\tau$                  &  $4.22 \pm 0.04$     &  $3.72 \pm 0.06$   
                                    && $4.37 \pm 0.05$     &  $3.80 \pm 0.06$    \\
         
         &  Cnorm                    &  $1.85 \pm 0.04$     &  $1.91 \pm 0.04$   
                                    && $1.77 \pm 0.03$     &  $1.91 \pm 0.04$    \\

GAUSSIAN\_1 &  $E_{\rm line}$ (keV)   &  \multicolumn{5}{c}{ [6.4] }\\
                                
         &  $\sigma$ (keV)         &  $1.00 \pm 0.17$    &  $1.25 \pm 0.13$
                                   && $1.13 \pm 0.16$    &  $1.29 \pm 0.15$     \\
                                
         &  Norm                   &  $0.03 \pm 0.01$    &  $0.05 \pm 0.01$
                                   && $0.04 \pm 0.01$    &  $0.05 \pm 0.01$      \\

GAUSSIAN\_2 &  $E_{\rm line}$ (keV)   &  $1.04 \pm 0.01$    &  $1.04 \pm 0.01$            
                                   && $1.04 \pm 0.01$    &  $1.04 \pm 0.01$       \\
                                
         &  $\sigma$ (keV)         &  $0.08 \pm 0.01$    &  $0.09 \pm 0.01$
                                   && $0.08 \pm 0.01$    &  $0.08 \pm 0.01$     \\
                                
         &  Norm                   &  $0.04 \pm 0.01$    &  $0.06 \pm 0.01$
                                   && $0.04 \pm 0.01$    &  $0.06 \pm 0.01$      \\
         
\hline
         &  $\chi^{2}/{\rm dof}$   &      \multicolumn{2}{c}{ 306.2/300 }
                                   &&     \multicolumn{2}{c}{ 293.7/299 }  \\ 
\hline
        &   $F_{\rm diskbb}$       &   $7.6 \pm 0.3$    &   $8.3 \pm 0.4$  
                                   &&  $6.9 \pm 0.3$    &   $7.8 \pm 0.3$                 \\
        &   $F_{\rm comptt}$       &   $22.2 \pm 0.3$   &   $22.3 \pm 0.4$
                                   &&  $21.2 \pm 0.3$   &   $22.3 \pm 0.3$              \\
        &   $F_{\rm gaussian\_1}$  &   $0.3 \pm 0.1$    &   $0.5 \pm 0.1$
                                   &&  $0.4 \pm 0.1$    &    $0.5 \pm 0.1$              \\
        &   $F_{\rm gaussian\_2}$  &   $0.07 \pm 0.01$    &   $0.10 \pm 0.01$
                                   &&   $0.07 \pm 0.01$   &   $0.09 \pm 0.01$               \\

\hline
\end{tabular}
\end{table*}

\begin{figure*}[ht]
\begin{center}
\resizebox{2\columnwidth}{!}{\rotatebox{0}{\includegraphics{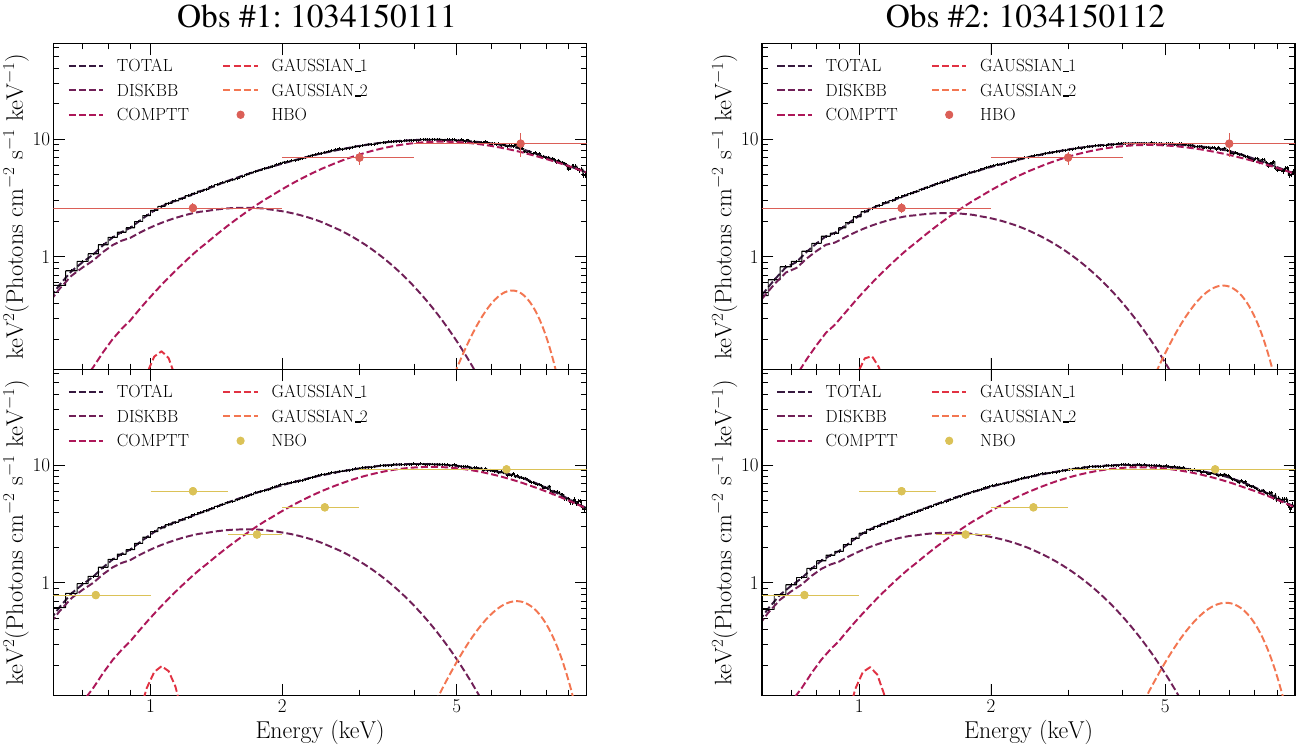}}}
\end{center}
\caption{Best-fitting spectral models for the spectra of the orbit with an HBO (upper panel) and an NBO (lower panel), together with the corresponding QPO spectra. The spectra were fitted with the model \texttt{tbfeo*(diskbb+comptt+gaussian+gaussian)}. Individual model components are marked with dotted lines in different colors. For better comparison, the HBO and NBO spectra are scaled by a factor of 400 and 200, respectively.}
\label{fig:fit}
\end{figure*}

For each observation, we fitted the HBO and NBO spectra jointly with a model consisting of a soft multicolor blackbody component (\texttt{diskbb}, \citealt{Mitsuda1984}) and a thermal Comptonization component (\texttt{comptt}, \citealt{Titarchuk1994}).
As the electron temperature, $kT_{\rm e}$, and optical depth, $\tau$, of the Comptonization component cannot be constrained simultaneously, we fixed $kT_{\rm e}=3$ keV, in line with previous results \citep[e.g.,][]{Disalvo2002, Done2002,Farinelli2009,Ludlam2022}.
\citet{Ludlam2022} found that $kT_{\rm e}$ does not change substantially during the transition from HB to NB.
Additionally, the \nicer\ spectra show clear evidence of a broadened Fe K line at around 6.4 keV and a low-energy emission line near 1 keV, both of which have been previously detected by other X-ray missions \citep[e.g.,][]{Disalvo2002,Ludlam2022}. These features were modeled using two \texttt{gaussian} lines, with the line energy of the Fe K line fixed at 6.4 keV due to poor constraint. Fixing the line energy at 6.7 keV \citep{Sudha2025} did not alter our primary findings; however, it resulted in a higher chi-square value.
The absorption along the line of sight was modeled using the component \texttt{tbfeo}\footnote{\url{https://pulsar.sternwarte.uni-erlangen.de/wilms/research/tbabs/}}, with the abundance tables from  \citet{Wilms2000} and the cross-section tables from \citet{Verner1996}. The column density, $N_{\rm H}$, and the abundance of oxygen, $A_{\rm O}$, in \texttt{tbfeo} were allowed to vary but were linked between the HBO and NBO spectra. All other parameters were allowed to vary independently between the two spectra.
The best-fitting spectral parameters are presented in Table \ref{tab:spectral_fits}. The unfolded spectra and model components are shown in Figure \ref{fig:fit}. 
In Figure \ref{fig:mcmc}, we show the distributions of the main spectral parameters.

The inner disc temperature, $kT_{\rm in}$, and the input seed photon temperature of the Comptonization component, $kT_{\rm 0}$, are consistent within errors between the HBO and NBO spectra.
Additionally, the disc normalization and the parameters of the \texttt{gaussian} components display only minor variations, with the contribution of the two emission lines to the total flux remaining below 2 per cent in both observations.  
A key difference we found during the transition is in the optical depth of the \texttt{comptt} component, $\tau$, which decreases significantly from the HBO epoch to the NBO epoch.

In order to compare the total time-averaged energy spectra to that of the fast variability, we converted the absolute rms amplitudes of both the HBO and NBO to the same units as the total spectra by considering the response of the instrument. This enabled us to construct QPO spectra for both the HBO and NBO. The resulting QPO spectra are shown in Figure \ref{fig:fit}.
It is clear that the shapes of the QPO spectra for both the HBO and NBO resemble that of the Comptonization component, suggesting that the BL/SL is driving the variability\footnote{Notice that the rms spectrum of a variable Comptonization component is not the same as the time-averaged spectrum of that same component (see, e.g., \citealt{Karpouzas2020})}.

\section{Discussion and conclusions}
\label{sec:discussion}

We have reported the detection of a rapid transition from a horizontal-branch oscillation to a normal-branch oscillation in two \nicer\ observations of \target. The large effective area of \nicer\ enables us to perform a detailed spectral-timing analysis of these transitions on short timescales, marking the first study to include spectral information in the 0.5--2 keV band.
We found that while the parameters of the disc component do not change significantly during the transitions from HBO to NBO, the optical depth of the Comptonization component decreases significantly assuming a fixed electron temperature.
Below we discuss our main results.

\subsection{A comparison with the transition from the hard-intermediate to the soft-intermediate state in BH-LMXBs}

A rapid switch from type-C to type-B QPOs is usually observed during the transition from the hard-intermediate to the soft-intermediate state in BH-LMXBs \citep[e.g.,][]{Homan2001,Soleri2008,Motta2011,Homan2020,Zhang2021,Yang2023}. The type-C to type-B QPO transitions are accompanied by an increase in the 0.5--10 keV flux and a spectral softening \citep{Homan2020,Yang2023}, which resemble the HBO to NBO transitions we observed in \target.
However, the frequency difference between HBO and NBO is greater than that observed between type-C and type-B QPOs. The type-C and type-B QPOs detected in the hard- and soft-intermediate states typically have a centroid frequency of less than 10 Hz \citep{Ingram2019}. 
In addition, the ratio between the spectrum with the type-B QPO and that with the type-C QPO (type-B/type-C) differs significantly from the ratio between the spectrum containing the NBO and the one with the HBO (NBO/HBO). 
\citet{Yang2023} found that the spectral ratio for type-B/type-C shows a peak around 4 keV and then gradually decreases with increasing photon energy, changing from above unity to below unity at around 10 keV.
This difference in spectral ratios is understandable, as the properties of the accretion flows vary between BH- and NS-LMXBs \citep{Munoz2014}. Notably, NS-LMXBs also exhibit emission from the surface of the neutron star.

The observed spectral-timing variation possibly suggests a change in the accretion geometry.
\citet{Ma2023} conducted a joint fit of the time-averaged spectra, as well as the rms and lag spectra of the QPOs, using the time-dependent Comptonization model \texttt{vkompth} \citep{Karpouzas2020,Bellavita2022} to investigate the geometry of the corona during the transition from type-C to type-B QPOs in the BH-LMXB MAXI J1820+070.
Taking the jet evolution into account, they proposed that the corona, which extends horizontally during the type-C QPO phase, transforms into a vertical, jet-like, structure during the type-B QPO phase.
In the same source, \citet{Wang2021} found that the characteristic reverberation lag frequency decreases with an increasing amplitude of the time lag during the transition, likely due to an expanding corona. 
In BH-LMXBs, the transition from type-C to type-B QPOs is typically accompanied by a significant change in jet properties. The steady compact jet switches off, and a large-scale transient jet is launched around the time the type-B QPOs appear \citep{Soleri2008,Fender2009,Russell2019,Homan2020,Carotenuto2021}. 
These findings suggest that the corona, potentially serving as the base of the jet, may undergo geometric changes that are closely coupled with the observed variations in jet properties during the transition \citep{Markoff2005,Mendez2022}. 
In NS-LMXBs, potential changes in jet properties have been observed in Scorpius X-1 during the HB to NB state transition \citep{Migliari2006}.
Combined with the observed significant spectral-timing evolution during the transition from HBO to NBO, these findings hint at a possible change in the accretion geometry.

\subsection{Possible changes in accretion geometry during the transition from HBO to NBO}

Our spectral analysis has shown that the parameters of the disc component do not vary significantly during the transitions from HBO to NBO.
Recently, \citet{Ludlam2022} performed an analysis of the reflection spectrum of \target\ using observations from \nicer\ and {\textit NuSTAR}. During those observations, the source traced out the HB, NB, and the vertex between the two. They found that the inner disc radius remains close to the NS ($\sim6.5 R_{\rm g}$), regardless of the source position along the Z-track.
Based on their and our findings, we conclude that the rapid QPO transitions observed in \target\ are primarily driven by changes in the Comptonization component rather than the disc.

The Comptonization component in NS-LMXBs is believed to originate from the hot BL/SL.
Using \textit{IXPE}, \citet{Farinelli2023} measured the X-ray polarization of \target\ when the source was in the upper NB to investigate the geometry of the BL/SL. A transient 5-Hz NBO was detected in simultaneous \nicer\ observations. The results of the spectro-polarimetric analysis provided a hint of a 90$^{\circ}$ rotation between the polarization angle of the disc and the Comptonization components. The Comptonization component exhibited a polarization degree $P = 4.0 \pm 0.7$ per cent with a polarization angle aligned with the radio jet. These findings strongly suggest that Comptonization takes place in a vertically extended BL/SL rather than in a coplanar region within the disc plane. However, the possibility of a significant contribution to the polarization signal from disc reflection cannot be excluded \citep{Farinelli2023}.
Our analysis reveals that both the HBO and NBO are likely linked to modulations in the BL/SL, as evidenced by the similarity in the shapes of their rms spectra to that of the Comptonization component. The spectral evolution during the transition from the HBO to the NBO epoch is characterized by a significant decrease in the optical depth of the Comptonization component, assuming a fixed electron temperature, accompanied by an increase in the source flux.
This drop in optical depth may be attributed to the expansion of the BL/SL, similar to the scenarios proposed to explain the hard- to soft-intermediate state transition in BH-LMXBs \citep{Wang2021,Ma2023}.

\subsection{Physical origins of the HBO and NBO}

Previous studies have demonstrated strong similarities in the variability properties of BH- and weakly magnetized NS-LMXBs \citep{Casella2005,Klein2008,Motta2017}. This suggests that the low-frequency QPOs in both types of systems may share a common physical origin. The energy-dependent properties of both type-C and type-B QPOs indicate that these QPOs are generated in the Comptonization region, which could be either the corona or the base of the jet \citep[e.g.,][]{Ma2021,Liu2022,Mendez2022}.
In our study, we found that the evolution of the fractional rms for the HBO and NBO, as obtained from \nicer, is consistent with findings from other missions \citep{Wijnads2001,Chhangte2022,Jia2023,Sudha2025}. For both the HBO and NBO, the shapes of their QPO spectra are similar to that of the Comptonization component, suggesting that the HBO and NBO may originate from flux modulation in the BL/SL.

The Lense-Thirring precession of the hot inner flow situated between a truncated radius, $r_{0}$, and the surface of the NS is a plausible model for explaining HBOs \citep{Stella1998,Ingram2010}. In this model, the QPO frequency increases as $r_{0}$ decreases. The 50-Hz HBO observed in \target\ during the transitions requires $r_{0}\sim 7 R_{\rm g}$, assuming a constant surface density for a 1.4 $M_{\odot}$ NS. This truncated radius estimated from the QPO frequency is in close agreement with the inner disc radius measured from the modeling of the reflection spectra \citep{Ludlam2022}. 
If NBOs originate from the same mechanism, the 5-Hz NBO observed during the transitions of \target\ would require $r_{0} \sim 20 R_{\rm g}$, which is inconsistent with the inner disc radius measured from the spectra of the NB \citep{Ludlam2022}.
Consequently, the Lense-Thirring precession of the hot inner flow cannot explain the NBO observed in our study.
\citet{Altamirano2012} also noted that the Lense-Thirring precession model fails to explain the 35--50 Hz QPOs observed in the 11\,Hz accreting pulsar IGRJ17480--2446 located in the globular cluster Terzan 5 (but see also \citealt{Buisson2021}).

NBOs could be oscillations in the optical depth of a radial inflow such as the BL/SL \citep{Fortner1989,Jia2023}. If the radial inflow begins approximately 300 km from the neutron star, the frequency of these oscillations would be around 5 Hz, which is comparable to the observed frequencies of NBOs \citep{Fortner1989}. However, the X-ray polarimetry results favor a vertically extended BL/SL rather than a radially extended one. Additionally, the size of the inflow estimated from the QPO frequency is much larger than the inner disc radius measured from the spectra \citep{Ludlam2022}.

One possible scenario is that both the HBO and NBO are the result of coupled oscillations between the Comptonization region and the accretion disc, as proposed in the model that explains the QPOs observed in BH-LMXBs \citep{Bellavita2022,Mastichiadis2022}. A similar mechanism could also be applicable to NS-LMXBs.
In this model, the QPO frequency could represent the characteristic frequency of a dynamical Comptonization region, influenced by factors such as the size of the region, the fraction of hard photons that return to the disc, the mass of the compact object, and the mass accretion rate \citep{Mastichiadis2022}. 
Assuming this scenario, variations of the BL/SL may lead to different types of QPOs seen in NS-LMXBs.

It is important to note that our \nicer\ spectral-timing results regarding the fast transitions do not provide strong constraints on the physical origin of the QPOs.
A systematic analysis of these transitions in other Z sources, especially through simultaneous spectral-timing-polarimetric studies on short timescales using future telescopes, such as the Enhanced X-ray Timing and Polarimetry Observatory (eXTP, \citealt{Zhang2019}), will aid in deepening our understanding of their origin.

\section{Acknowledgments}
\begin{acknowledgments}
This work was supported by the National Natural Science Foundation of China (NSFC) under grants 12203052 and 12403053.
M.M. acknowledges the research program Athena with project No.184.034.002, which is (partly) financed by the Dutch Research Council (NWO). 
This work was supported by NASA through the \nicer\ mission and the Astrophysics Explorers Program, and made use of data and software provided by the High  Energy Astrophysics Science Archive Research Center (HEASARC).
\end{acknowledgments}

%






\appendix

\begin{figure*}[ht]
\begin{center}
\resizebox{\columnwidth}{!}{\rotatebox{0}{\includegraphics{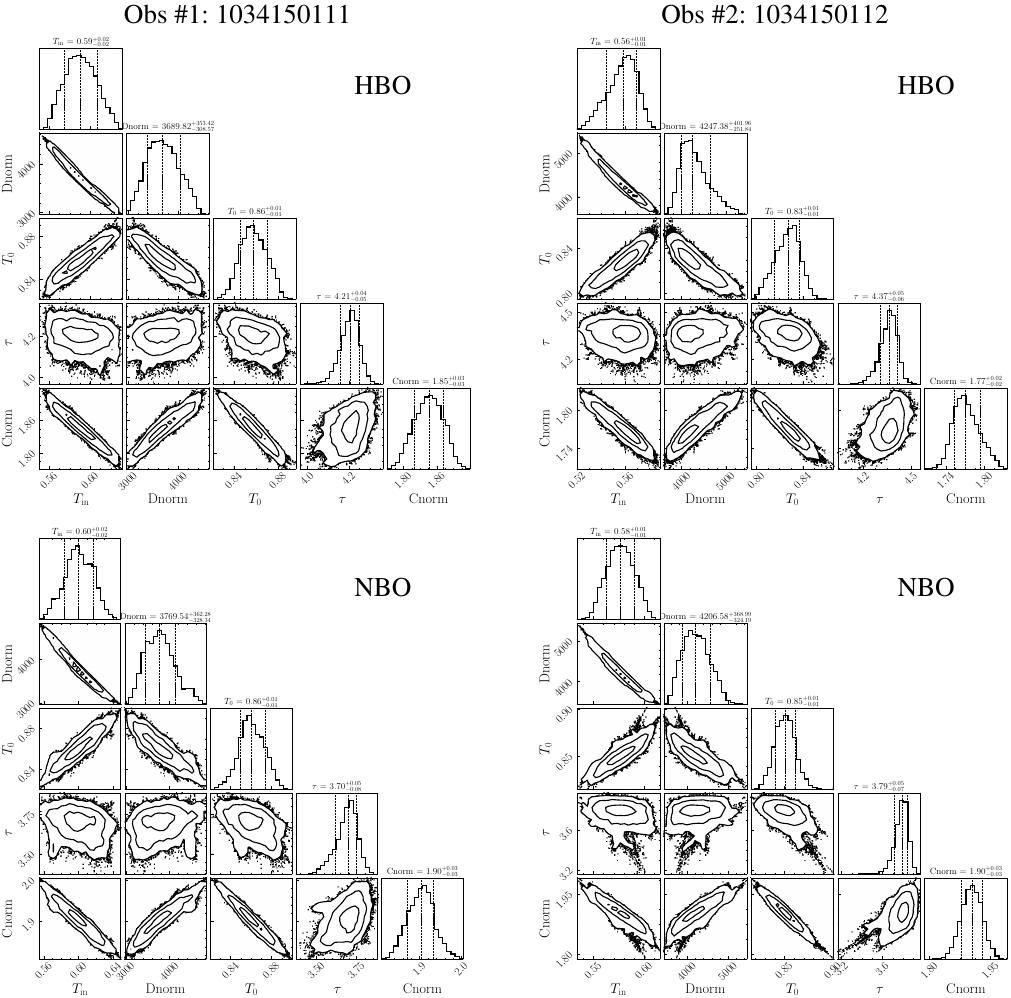}}}
\end{center}
\caption{Distributions of the main spectral parameters obtained from the MCMC analysis. The left two panels display the results from the HBO (upper) and NBO (lower) epochs for Obs \#1, while the right two panels show the corresponding results from the HBO (upper) and NBO (lower) epochs for Obs \#2. The spectra were fitted with the model \texttt{tbfeo*(diskbb+comptt+gaussian+gaussian)}. The spectral parameters shown are the disc temperature ($T_{\rm in}$), the normalization of the disc component (Dnorm), the input seed photon temperature of the Comptonization component ($T_{0}$), the optical depth ($\tau$), and the normalization of the Comptonization component (Cnorm). The contours in the plots correspond to confidence levels of  1$\sigma$, 2$\sigma$, and 3$\sigma$.}
\label{fig:mcmc}
\end{figure*}

\end{document}